\documentclass[prl,twocolumn,english,superscriptaddress,floatfix,longbibliography]{revtex4-2}

\usepackage{ulem}
\usepackage{graphicx}% Include figure files
\usepackage{dcolumn}% Align table columns on decimal point
\usepackage{bm}% bold math
\usepackage{xcolor}
\usepackage{amsmath}
\usepackage{amssymb}
\usepackage{lineno}

\usepackage{hyperref}
\hypersetup{colorlinks=true, citecolor=blue, linkcolor=black, urlcolor=blue}

\newcommand{\ket}[1]{|#1\rangle}
\newcommand{\bra}[1]{\langle#1|}

\newcommand{\ketbra}[2]{|#1\rangle\langle #2|}

\begin{document}
%\linenumbers

%\title{Non-Hermitian Jarzynski equality}
%\title{Work fluctuations in the non-Hermitian quantum dynamics of a superconducting qubit}
%\title{Constraining work fluctuations of non-Hermitian dynamics across the exceptional point of a dissipative qubit}
\title{Constraining work fluctuations of non-Hermitian dynamics across the exceptional point of a superconducting qubit}

\author{Serra Erdamar}
\affiliation{Department of Electrical and Systems Engineering, Washington University, St. Louis, Missouri 63130}
\affiliation{Department of Physics, Washington University, St. Louis, Missouri 63130}
\author{Maryam Abbasi}
\author{Byung Ha}
\author{Weijian Chen}
\affiliation{Department of Physics, Washington University, St. Louis, Missouri 63130}
\author{Jacob Muldoon}
\author{Yogesh Joglekar}
\affiliation{Department of Physics, Indiana University-Purdue University Indianapolis, Indianapolis, Indiana 46202}
\author{Kater W. Murch}
\affiliation{Department of Physics, Washington University, St. Louis, Missouri 63130}

\date{\today}% It is always \today, today,

%=========================================================================================%
             
\begin{abstract}
%thanks chatGPT!
%\linenumbers
%The thermodynamics of small quantum systems is distinguished by fluctuating work, 
Thermodynamics constrains changes to the energy of a system, both deliberate and random, via its first and second laws. When the system is not in equilibrium, fluctuation theorems such as the Jarzynski equality further restrict the distributions of deliberate work done. Such fluctuation theorems have been experimentally verified in small, non-equilibrium quantum systems undergoing unitary or decohering dynamics. Yet, their validity in systems governed by a non-Hermitian Hamiltonian has long been contentious, due to the false premise of the Hamiltonian's dual and equivalent roles in dynamics and energetics. Here we show that work fluctuations in a non-Hermitian qubit obey the Jarzynski equality even if its Hamiltonian has complex or purely imaginary eigenvalues. With post-selection on a dissipative superconducting circuit undergoing a cyclic parameter sweep, we experimentally quantify the work distribution using projective energy measurements and show that the fate of the Jarzynski equality is determined by the parity-time symmetry of, and the energetics that result from, the corresponding non-Hermitian, Floquet Hamiltonian.  
By distinguishing the energetics from non-Hermitian dynamics, our results provide the recipe for investigating the non-equilibrium quantum thermodynamics of such open systems.  

%In quantum systems, work fluctuations dominate\serracom{*} the dynamics of the process. By utilizing the two-time projective measurement protocol, the work can be characterized by a difference in eigenenergy. To better understand the thermodynamics of quantum systems, we relate the obtained work distribution to the difference in free energy through the Jarzynski equality. Specifically, we study the dynamics of a non-Hermitian system realized on a sub-manifold of a superconducting transmon qubit. The dissipation is engineered such that post selecting on this manifold provides\serracom{*} an effective non-Hermitian Hamiltonian. Here, we find that the complex energy spectrum affects the validity of the Jarzynski equality. Since our non-Hermitian Hamiltonian obeys PT symmetry, the complexity of the energy spectrum can be compartmentalized\serracom{*} in the Hamiltonian’s parameter space. We show that the Jarzynski equality remains valid in regions of unbroken PT-symmetry. However, in regions of broken PT-symmetry, the complex energy spectrum results in violation of the Jarzynski equality. Characterization of this violation forms the basis for an extension of the Jarzynski equality for non-Hermitian dynamics, where complex energy evolution can be related to changes in information.
\end{abstract}

\maketitle
%=========================================================================================%
%\linenumbers

\begin{figure}[t]
    % \centering
   \includegraphics[width=1\linewidth]{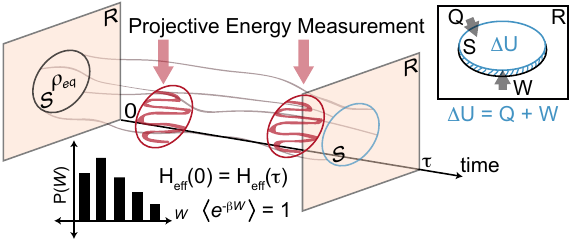}
   %\internallinenumbers
    \caption{{\bf Thermodynamics of open quantum systems.}  The first law of thermodynamics (right inset) states that the internal energy of a system (S) coupled to a reservoir (R) is additively changed by the heat $Q$ and the work $W$. The generalized second law or the Jarzynski equality (\ref{eq:JE1}) governs the trajectory-dependent work fluctuations. For a quantum system starting in an equilibrium density matrix $\rho_\textrm{eq}$, these work fluctuations are characterized by projective energy measurements leading to a discrete work distribution $P(W)$. We show that for cyclic parameter variations in $H_\mathrm{eff}(t)$, the average exponentiated work is unity when the corresponding Floquet Hamiltonian $H_\mathrm{eff}^F$ has a parity-time symmetry, and its Floquet energy operator matches the system's initial energy operator.}
    \label{introconcept_v2}
\end{figure}

\noindent{\bf Introduction}. The concept of a small system coupled to a large reservoir is elemental to both thermodynamics and open quantum systems. In thermodynamics, a reservoir allows one to distinguish between two types of energetics: heat $Q$, the random energy transferred to the system from the reservoir, and work $W$, deliberately imparted to the system. The energy $U$ of the system is additively changed by the two, thereby encoding the first law of quantum thermodynamics, $\Delta U=Q+W$ (Fig.~\ref{introconcept_v2} right inset)~\cite{Ali79,Kos84,Spohn2007,Hor12,alonso16,Elo17}. Conversely, a closed quantum system is governed by a Hermitian Hamiltonian $H(t)$, undergoes unitary evolution with zero heat exchange, and its energy is equal to the expectation value of the generator of its dynamics $H(t)$.  %froreservoir introduces the concept of (heat) energy transfer between systems, contrasted with energy imparted directly on the system in the form of work. provides the notion of temperature, and is therefore important to formulations of the second law. Fluctuation theorems utilize equalities that relate fluctuations of, for example, work, to equilibrium thermodynamics properties. %which are now \Maryam{\sout{now}} commonly stated as fluctuation theorems---equalities that relate fluctuations of, for example, work to equilibrium thermodynamical properties.% 
When coupled to a reservoir, one describes its evolution by averaging over possible, consistent micro-states of the reservoir. This averaging leads to (engineered) decoherence and dissipation~\cite{Harrington2022}; 
%~\cite{Mur12, Har19, Har22,Ver09}, 
%\st{which are useful for quantum control but deleterious to coherent quantum evolution}~\cite{Mur12, Har19, Har22,Ver09}. Under the Markov approximation, 
the resulting dynamics are described by a Lindblad equation $\partial_t\rho=\mathcal{L}\rho$ for the reduced density matrix $\rho(t)$ of the system~\cite{lidar2020lecture}. Here, the system-reservoir coupling results in trajectory-dependent heat and work that, when added together, gives trajectory-independent change in the energy of the system~\cite{nagh20}. In such cases, the internal energy operator $H(t)$, which encodes the energy $U(t)\equiv\mathrm{Tr}[\rho(t)H(t)]$, is distinct from the generator ${\mathcal L}$ of its temporal dynamics. 

In addition to the work-energy theorem, work fluctuations of a non-equilibrium system with internal energy operator $H(t)$
%, coupled to reservoir, 
are further constrained by the Jarzynski equality~\cite{Jarzynski1997,Kurchan1998,Crooks1998,Tasaki2000}
\begin{equation}
    \label{eq:JE1}
    \langle e^{-\beta W}\rangle = \frac{Z(\tau)}{Z(0)}\equiv e^{-\beta\Delta F}.
\end{equation}
Here $\langle\cdot\rangle$ denotes trajectory-ensemble average, $\beta^{-1}$ is the reservoir temperature, $Z(t)\equiv\mathrm{Tr}\exp[-\beta H(t)]$ is the system partition function, and $\Delta F\equiv F(\tau)-F(0)$ is the Helmholtz free energy change in time $\tau$. The equality (\ref{eq:JE1}) supersedes the Jensen inequality $\Delta F\leq\langle W\rangle$ that constrains the amount of work done on a system and its free-energy change. In a quantum system with indefinite energy, a two-point-measurement (TPM) protocol quantifies changes in a system's energy $\Delta U$~\cite{Hub08}. It entails performing a pair of projective measurements in the energy basis to quantify $\Delta U$ in terms of transition probabilities between instantaneous eigenstates of the internal energy operator $H(t)$ (Fig.~\ref{introconcept_v2}). These transition probabilities differ for unitary and Lindblad evolution and yet Jarzynski equality (\ref{eq:JE1}) holds~\cite{Mukamel2003} for unital quantum maps~\cite{alba2013}, as has been experimentally verified~\cite{Bat14,An14,Cer17,Xio18,Smi18}. 

%When Eq.~\eqref{eq:JE1} that account for effects of measurement and feedback \cite{Sag10, Kos14,Nagh18,song21,Mas18}, as well as contributions of heat \cite{Pek15} have been investigated.

%===================== FIG 1 ========================================%
%===================== FIG 1 ========================================%
%\resetlinenumber[69]
In recent years, a third model of quantum dynamics obtained by post-selecting on quantum trajectories with no quantum jumps has emerged~\cite{nagh19,Abb22}. With a non-Hermitian generator $H_\mathrm{eff}(t)=H(t)+i\Gamma(t)$ and a nonlinear, trace-preserving equation of motion~\cite{Brody2012}, it maps pure states into pure states but changes the entropy of mixed states~\cite{Bian2020}, thereby commingling salient features of unitary and Lindblad evolution. When the non-Hermitian $H_\mathrm{eff}$ has a real spectrum, its role in dynamics has been conflated with energetics, leading to predicted violations of the Jarzynski equality and Crooks fluctuation theorem when the spectrum of $H_\mathrm{eff}$ turns complex~\cite{Def15,Gardas2016,Zeng2017,Bobo2018,Boboa2018,Zhou2021}. Fundamentally, the coherent, non-unitary, non-unital dynamics generated by $H_\mathrm{eff}$ begs the question: What are the constraints on quantum work fluctuations in such dynamics?

Here, we demonstrate the Jarzynski equality in a non-Hermitian qubit undergoing cyclic parameter changes, including cases where $H_\mathrm{eff}(t)$ has complex eigenvalues at all times. The qubit dynamics is characterized by the non-unitary $G(\tau)=\mathbb{T}\exp\left[-i\int_0^\tau H_\mathrm{eff}(t')dt'\right]\equiv\exp(-i\tau H^{F}_{\mathrm{eff}})$ that defines the non-Hermitian Floquet Hamiltonian $H^F_{\mathrm{eff}}\equiv H^F+i\Gamma^F$, where $H^F$ is the Floquet internal energy operator. The Jarzynski equality $\langle e^{-\beta W}\rangle=1$ is satisfied when $H^F_\mathrm{eff}$ has an explicit or emergent parity-time symmetry that guarantees real or complex-conjugate eigenvalues, and $H^F\propto H(0)$, i.e. the two internal energy operators have the same eigenbasis (see Methods). 

\begin{figure}[t]
    % \centering
    \includegraphics[width=1\linewidth]{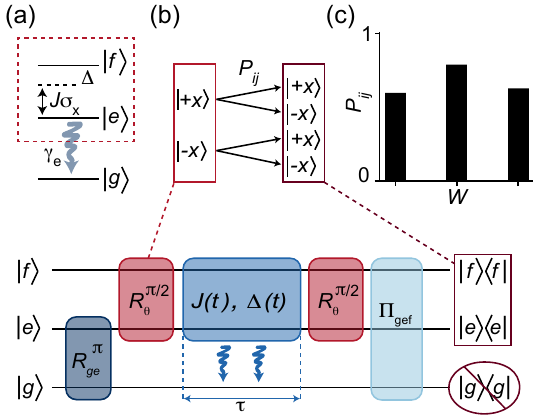}
    %\internallinenumbers
    \caption{{\bf Experimental setup}. (a) A non-Hermitian qubit is realized as the sub-manifold (dashed box) of the lowest three levels of a transmon circuit. The system exhibits decay from the $\ket{e}$ state to $\ket{g}$ at rate $\gamma_e$ and is driven by a microwave drive with detuning $\Delta(t)$ and coupling rate $J(t)$. (b) The experimental protocol involves preparing an initial eigenstate $\ket{+x}$ or $\ket{-x}$, dynamically tuning the energy operator $H(t)$ for a certain time $\tau$ and returning it to its initial value $H(\tau)=H(0)$, followed by post-selective quantum state tomography. (c) By sampling both initial states $\ket{j}\in \{\ket{+x},\ket{-x}\}$ and post-selecting cases where the system does not project onto the ground state, we determine the transition probabilities $P_{ij}$ within the excited-state manifold $\{\ket{e},\ket{f}\}$.}% b). This sequence is completed with another projective measurement to find the qubit in $|+\rangle$ or $|-\rangle$.} (b) In the parameter space defined by $J$ and $\Delta$ there is an exceptional point degeneracy (EP) at $J = J_\mathrm{EP}$ and $\Delta = 0$. For $J>J_\mathrm{EP}$ and $\Delta=0$ the eigenvalue difference is strictly real, corresponding to the region of unbroken PT symmetry. (c)
    %this might be too long and repetitive for what is in the text
    \label{Schematic}
\end{figure}
%===================== FIG 2 ========================================%
%\resetlinenumber[97]
\noindent {\bf Measuring work distribution.} Our experimental platform comprises a superconducting transmon circuit with energy eigenstates labeled $\{\ket{g}, \ket{e}, \ket{f}\}$ dispersively coupled to a microwave cavity (see Methods). Bath engineering allows us to tune the radiative decay rates such that the decay rate $\gamma_e = 1.57\ \mu\mathrm{s}^{-1}$ of state $\ket{e}$ by quantum jumps $\ket{e}\to\ket{g}$ is much larger than the decay rate $\gamma_f = 0.21\ \mu\mathrm{s}^{-1}$ leading to a decay contrast $\gamma\equiv \gamma_e-\gamma_f\approx \gamma_e$. 
%There is an additional decay channel weakly coupled to the qubit at our target frequency, $\omega_{ge} = 4.25$ GHz, leading to rapid radiative decay of the state $\ket{e}$ by quantum jumps $\ket{e}\to\ket{g}$ at rate $\gamma_e = 1.57\ \mu\mathrm{s}^{-1}$. The density of states of the environment is shaped such that $\gamma_e \gg \gamma_f$, with $\gamma_f = 0.21\ \mu\mathrm{s}^{-1}$ leading to a decay contrast $\gamma = \gamma_e-\gamma_f$. 
We post-select quantum trajectories with no quantum jumps to the $\ket{g}$ state, thereby limiting the dynamics only to the excited-state subspace $\{\ket{e}, \ket{f}\}$.
%from the dynamics by only considering experimental protocols that preserve the excited qubit subspace  $\{\ket{e}, \ket{f}\}$. 
With the addition of a drive that couples the  states $\ket{e}$ and $\ket{f}$ with detuning $\Delta(t)$ and rate $J(t)$ (Fig.~\ref{Schematic}a), the evolution of this qubit subspace is described by an effective non-Hermitian Hamiltonian, 
%(in a frame rotating with the drive)
\begin{equation} \label{H}
H_\mathrm{eff}(t)=J(t)\sigma_x+\Delta(t)\ket{f}\bra{f}+\frac{i\gamma}{4}\sigma_z=H(t)+i\Gamma,
%    H = \begin{pmatrix} \Delta - \frac{i\gamma}{4} & J \\ J & +\frac{i\gamma}{4} \end{pmatrix} = H_\mathrm{H} + H_\mathrm{AH}.
\end{equation}
where $\sigma_x=(\ket{f}\bra{e}+\ket{e}\bra{f})$ and $\sigma_z=(\ket{f}\bra{f}-\ket{e}\bra{e})$ are Pauli matrices in the excited-state subspace. 
%Here, we indicate that we can separate $H$ into Hermitian ($H_\mathrm{H}$) and anti-Hermitian ($H_\mathrm{AH}$) parts, respectively. Although the experimental system only consists of lossy dynamics, by post-selecting on the no-jump evolution we essentially re-scale the dynamics so that the state $\ket{f}$ has relative gain compared to state $\ket{e}$, giving the Hamiltonian a PT-symmetric form when $\Delta=0$. The phase of the drive defines the axis of the $\sigma_x\equiv \ketbra{f}{e}+ \ketbra{e}{f}$ Pauli operator. 
When $\Delta(t)\equiv 0$, the Hamiltonian $H_\mathrm{eff}(t)$ commutes with the antilinear operator $\mathcal{PT}=\sigma_x\mathcal{K}$, where $\mathcal{K}$ denotes complex conjugation, at all times. In the static case, this explicit $\mathcal{PT}$-symmetry underlies the purely real or purely imaginary eigenvalues $\lambda_{\pm}=\pm\sqrt{J^2-(\gamma/4)^2}$ of $H_\mathrm{eff}$ with an exceptional-point (EP) degeneracy at $J_\mathrm{EP}=\gamma/4$. For sampling the work distribution, we implement three time-periodic parameter paths, 
\begin{linenomath}
\begin{align}
    \label{Jvary}
    & J(t)=\Bar{J}+\frac{(J_{\max}-J_{\min})}{2}\cos\left(\frac{2\pi t}{\tau}\right),\,\, \Delta(t)=0\\
%    &J(t)=\frac{(J_{\max}+J_{\min})}{2}+\frac{(J_{\max}-J_{\min})}{2}
%    \cos\left(\frac{2\pi t}{\tau}\right), \\
    \label{Delta1vary}
    & \Delta_1(t)= \Delta_\mathrm{max}\sin\left(\frac{\pi t}{\tau}\right),\,\,J(t)=J_{\max}\\
    \label{Delta2vary}
    & \Delta_2(t)= \Delta_\mathrm{max}\sin\left(\frac{2\pi t}{\tau}\right),\,\,J(t)=J_{\max},
\end{align}
\end{linenomath}
where $\tau$ is the protocol duration and $\Bar{J}=(J_{\max}+J_{\min})/2$. When $\Delta(t)\neq 0$, the instantaneous eigenvalues $\lambda_\pm(t)$ of the non-Hermitian $H_\mathrm{eff}(t)$ are complex and the Hamiltonian has no explicit $\mathcal{PT}$-symmetry. 
The basis of the experiment is to determine the work distribution $P(W)$  after the system is driven by the cyclic internal energy operator $H(t)$ (Fig.~\ref{Schematic}b). In our experiment, the TPM procedure consists of three steps: (i)~With a sequence of resonant rotations to the transmon circuit, we initialize the system in the eigenstates $\ket{\pm x}=(\ket{f}\pm\ket{e})/\sqrt{2}$ of the energy operator $H(0)=J_{\max}\sigma_x$. %Such that $H_{J,\Delta}\ket{\pm} = \lambda_\pm \ket{\pm}$ for a target value of $J$ and $\Delta$.
A Gibbs state with inverse temperature $\beta$ is then synthesized by preparing the two eigenstates with relative probabilities $P_{\pm x}\propto\exp(\mp\beta J_{\max})$. 
%By preparing either eigenstate of $H_\mathrm{H}$ with probability  $P_\pm = \exp(\mp\beta J)/Z$, with $Z = \sum_\pm \exp(\mp\beta J)$, we synthesize an initial Gibbs state with inverse temperature $\beta$. 
Throughout this work we set %$\beta = 7.6 \times 10^8$ eV$^{-1}$ 
$\beta = 0.5 \ \mu\mathrm{s}/\mathrm{rad}$, which corresponds to $P_{+x}=0.98$. % normalized with the eigenvalue energy so that we can work with a unit-less constant. 
%We then abruptly apply a drive to realize the Hamiltonian $H_\mathrm{eff}$. 
(ii)~We dynamically apply work to the qubit by tuning the parameters $J(t), \Delta(t)$ as in Eqs.~(\ref{Jvary})-(\ref{Delta2vary}). (iii)~We perform a final projective measurement in the basis $\{\ket{g},\ket{+x},\ket{-x}\}$ via a single shot, multi-state readout of the qutrit, which gives probabilities $\{p_{g,j},p_{+x,j},p_{-x,j}\}$ that add up to unity. 
%For this, we abruptly switch $J=0$ and apply a resonant rotation to rotate the state $\ket{+x}$ ($\ket{-x}$) to the qutrit state $\ket{e}$ ($\ket{f}$) and perform a single shot, multi-state readout of the qutrit. %This measurement corresponds to three possible projectors $\{\ketbra{g}{g}, \ketbra{e}{e}, \ketbra{f}{f}\}$ (Fig.~\ref{Schematic}c light blue pulse). 
%The measurement is performed with moderate fidelities in the range of 50--97\% depending on the state mapping. We independently calibrate the measurement fidelities and appropriately scale the measured probabilities to determine fidelity-corrected probabilities $\{p_{+x,j}, p_{-x,j}, p_{g,j}\}$ with the system initialized in state $\ket{j}=\ket{\pm x}$. We then apply the postselection by rescaling the probabilities $P_{\pm x,j}= p_{\pm x,j}/(p_{+x,j}+ p_{-x,j})$.   %Th   Due to measurement infidelity for each state, we scale the readout using a matrix that is obtained from a $0$, $\pi_{ge}$, $\pi_{ef}$ pulse sequence. We then discard all trials resulting in $\ketbra{g}{g}$, leaving $N$ post-selected experimental trials. Of which we determine $P_{+j} = N_{+,\ketbra{e}{e}}/N$ and $P_{-j} = N_{-,\ketbra{f}{f}}/N$ based on the number of readouts resulting in $\ketbra{e}{e}$ and $\ketbra{f}{f}$.% \kater{[Let's rewrite above paragraph to reflect what we actually do, we rescale and adjust for measurement fidelity]}
The TPM protocol determines the total energy change $\Delta U=W+Q$ whose distribution is characterized by the transition probabilities~\cite{Varma2023} 
\begin{linenomath}
\begin{align}
    \label{eq:pij}
    P_{ij}(\tau)=\frac{|\bra{i}G(\tau)\ket{j}|^2}{\bra{j}G^\dagger(\tau)G(\tau)\ket{j}}=\frac{p_{ij}}{p_{+x,j}+p_{-x,j}}
\end{align}
\end{linenomath}
 where $i,j=\pm x$ label the eigenstates of the internal energy operator $H(0)$. The state-dependent denominator in \eqref{eq:pij} captures the norm-preserving nature of the post-selection process, $\sum_i P_{ij}(\tau)=1$ (Fig.~\ref{Schematic}b,c). The exponentiated-work expectation value \eqref{eq:JE1} is obtained as
\begin{equation}
     \langle e^{-\beta W}\rangle(\tau) = \sum_{ij=\pm x} e^{-\beta J_{\max}(i-j)} P_{ij}(\tau)P_{j},
     %\tilde{P}_{ij}e^{-\beta W_{ij}}.%= P_+P_{++}e^{-\beta W_{++}}+\\P_+P_{+-}e^{-\beta W_{+-}}+P_-P_{--}e^{-\beta W_{--}}+P_-P_{-+}e^{-\beta W_{-+}},
     \label{LHS}
\end{equation}
where the statistical weights $P_{\pm x}=\{0.98,0.02\}$ reflect the reservoir temperature and transition probabilities $P_{ij}(\tau)$ are experimentally measured for loop duration $\tau$ ranging from $0.1$ $\mu\mathrm{s} \leq\tau\leq 1$ $\mu\mathrm{s}$. 
%\kater{Here, $P_i$ is determined by the sampling distribution of initial states prepared in Step 1, and $P_i$, and $P_{ij}$ are the scaled transition probabilities associated with $H_{NH}$.}

%===================== FIG 3 ========================================%
\begin{figure}[t]
    % \centering
    \includegraphics[width=1\linewidth]{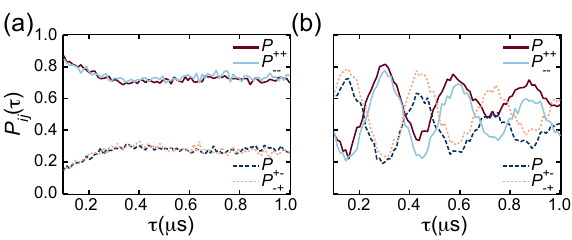}
    %\internallinenumbers
    \caption{{\bf Symmetries of Transition probabilities}. (a) Measured probabilities $P_{ij}(\tau)$ for the first path, \eqref{Jvary}, with $J_\mathrm{max}=J_\mathrm{min} = 3.74\ \mathrm{rad}/ \mu\mathrm{s}$ are symmetrical under the exchange $+x\leftrightarrow -x$. (b) Measured probabilities for the second path, \eqref{Delta1vary}, with the same $J_{\max}$ value and $\Delta_{\max} = 10\pi\ \mathrm{rad}/\mu\mathrm{s}$ show clear asymmetry, $P_{++}\neq P_{--}$ and $P_{+-}\neq P_{-+}$. This asymmetry is connected to the absence of $\mathcal{PT}$-symmetry for the Hamiltonian $H_\mathrm{eff}(t)$ along the second cyclic path.}    
    %$\Delta_\mathrm{max}/2\pi = 5 \mathrm{MHz}$
    \label{Probvstime}
\end{figure}
%===================== FIG 3 ========================================%
 
%To measure the probability that the qubit ends in the final state i and started in an initial state j, we first compute the initial state probability given in Eq.\eqref{initialProb}. Next, we find the conditional probability of ending in state i given the qubit started in state j. In experiment, the conditional probability $P_{i|j}$, is obtained from performing the projective measurements from step 3 $N \gg 1$ times and obtaining a ratio of landing in the state i. The projective measurement consists of projecting the qubit state onto a specific basis: $|\pm\rangle$. 
%\begin{equation}
%\label{conditionalprob}
%    P_{ij}=P(j\cap i) = P_jP_{i|j}
%\end{equation}
%We note that $J(t = 0) = J(t = \tau) = J_{max}$, which implies that the Hamiltonian at the end of the process is equivalent to the initial Hamiltonian. This simplifies our calculation of $e^{-\beta \Delta F}$ in the Jarzynski equality since $\Delta F = 0$, and, consequently, $e^{-\beta \Delta F} = 1$. 

%For each parameter path, Eqs.~(\ref{Jvary})-(\ref{Delta2vary}), we measure $P_{ij}(\tau)$ for loop duration $\tau$ ranging from $100$ ns to $1$ $\mu \mathrm{s}$. 
%\resetlinenumber[154]
Figure~\ref{Probvstime}a shows that for the first path with zero detuning, Eq.~(\ref{Jvary}), the survival and transition probabilities are equal for the two energy eigenstates. On the contrary, for the second path, Eq.~(\ref{Delta1vary}) with $\Delta_{\max} = 10\pi\ \mathrm{rad}/\mu\mathrm{s}$, the probabilities $P_{++}\neq P_{--}$ (or equivalently, $P_{-+}\neq P_{+-}$) are clearly asymmetrical (Fig.~\ref{Probvstime}b). Both cases have $J_{\mathrm{max}} = J_{\mathrm{min}} = 3.74\ \mathrm{rad}/ \mu\mathrm{s}$. We observe stark differences between these two cases which correspond to Hamiltonians $H_\mathrm{eff}(t)$ with or without an explicit $\mathcal{PT}$ symmetry, respectively. 
%With $\Delta=0$, transition probabilities that end in $\ket{\pm x}$ are equal, corresponding to a unital (identity preserving) map, whereas with $\Delta_\mathrm{max}\neq0$  the transition probabilities acquire an asymmetry. 

%Fig.~\ref{Probvstime}. Here we choose a parameter path with fixed  %Where maroon $P_{++}$ indicates a transition from $\ket{+x}$ to $\ket{+x}$, light blue $P_{--}$ indicates a transition from $\ket{-x}$ to $\ket{-x}$, dark blue $P_{+-}$ indicates a transition from $\ket{-x}$ to $\ket{+x}$, and orange $P_{-+}$ indicates a transition from $\ket{+x}$ to $\ket{-x}$. 
% $J_{\mathrm{max}} = J_{\mathrm{min}} = 3.74\ \mathrm{rad}/ \mu\mathrm{s}$ and either fixed $\Delta=0$ (Fig.~\ref{Probvstime}a) or $\Delta_\mathrm{max} = 10\pi\ \mathrm{rad}/\mu\mathrm{s}$ (Fig.~\ref{Probvstime}b). 

%=========================================================================================%

%===================== FIG 4 ========================================%
 
\begin{figure}[t]
    \includegraphics[width=1\linewidth]{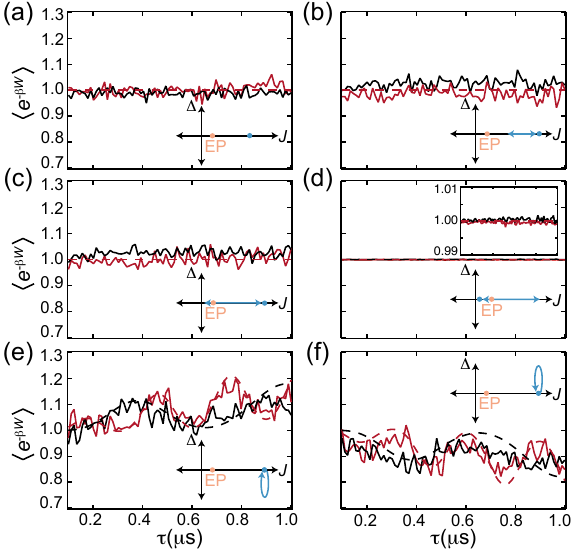}
    %\internallinenumbers
   \caption{{\bf Jarzynski equality and its violation.} 
   Experimental calculation of $\langle e^{-\beta W}\rangle(\tau)$ (solid lines) alongside the simulation results (dashed lines). The insets depict the parameter paths $J(t)$ or $\Delta_1(t)$, and the EP at $J_\mathrm{EP}=\gamma/4$.  (a-d) $J(t)$ sweeps with $J_{\mathrm{max}} = 3.74$ rad/$\mu \mathrm{s}$ (red), $J_{\mathrm{max}} = 1.89$ rad/$\mu \mathrm{s}$ (black) (a) The static case, $J_{\max}=J_{\min}$ satisfies \eqref{eq:JE1}. (b) The path going from $J_{\max}$ to $J_{\min}=0.5J_{\max}$ is in the $\mathcal{PT}$-symmetric region and satisfies \eqref{eq:JE1}, as does the path (c) going across the EP to $J_{\min}=0$. (d) For a path starting in the $\mathcal{PT}$-broken region at $J_{\max}=0.04$ rad/$\mu \mathrm{s}$ and reaching across the EP to $J_{\min}>J_{\max}$, the Jarzynski equality \eqref{eq:JE1} holds (upper inset: zoomed-in view of fluctuations). (e-f) For $\Delta_1(t)$ sweeps with $J_{\max}=3.74$ rad/$\mu \mathrm{s}$, the average exponentiated-work increasingly deviates from unity: $\Delta_{\max}=\mp 10\pi$ rad/$\mu \mathrm{s}$ (red), $\Delta_{\max}=\mp 2\pi$ rad/$\mu \mathrm{s}$ (black). The absence of transition probability symmetry seen in Fig.~\ref{Probvstime}b (or a parity-time symmetry) is instrumental to the violation of the Jarzynski equality.}
    \label{Results}
\end{figure}
%===================== FIG 4 ========================================%

%\resetlinenumber[164]
\noindent{\bf Jarzynski equality and its violation.} Figure \ref{Results} summarizes the experimental results for $\langle e^{-\beta W}\rangle(\tau)$ for $J(t)$ variations (a-d) and $\Delta_1(t)$ variations (e-f); each parameter path and the location of the EP is schematically shown in the corresponding panel inset. We see that $\langle e^{-\beta W}\rangle\simeq 1$ for $J(t)$ variations that range from the static case, $J_{\max}=J_{\min}$ (a), to paths confined to the $\mathcal{PT}$-symmetric region, $J_{\min}=0.5J_{\max}$ (b), to paths that traverse across the EP into the $\mathcal{PT}$-broken region with $J_{\min}=0$ (c). Panel (d) shows that starting from the $\mathcal{PT}$-broken region and traversing across the EP into the $\mathcal{PT}$-symmetric region also maintains the Jarzynski equality, with much smaller fluctuations arising from the smaller energy scale (upper inset). Thus, the Jarzynski equality is satisfied for arbitrary $J(t)$ sweeps, independent of the real or imaginary nature of eigenvalues of $H_\mathrm{eff}(t)$ as long as the Hamiltonian has an explicit $\mathcal{PT}$-symmetry. For the first path, this explicit $\mathcal{PT}$-symmetry also ensures that the corresponding Floquet energy operator $H^F$ has the same energetics as the system's initial energy operator $H(0)=J_{\max}\sigma_x$. In sharp contrast, for one-sided sweeps $\Delta_1(t)$, the average exponentiated-work $\langle e^{-\beta W}\rangle(\tau)$ exceeds one for $\Delta_{\max}<0$ (e) and is below unity for $\Delta_{\max}>0$ (f), thereby indicating that the Jarzynski equality is violated when $H_{\mathrm{eff}}(t)$ or its Floquet counterpart $H^F_\mathrm{eff}$ do not have an antilinear (parity-time) symmetry. 

%. We consider six potential parameter paths, as indicated by insets in each panel. In Fig.~\ref{Results}a-d the red lines display $\langle e^{-\beta W}\rangle$ for $J_\mathrm{max} = 3.74\ \mathrm{rad}/ \mu\mathrm{s}$ and the black lines for $J_\mathrm{max} = 1.89\ \mathrm{rad}/ \mu\mathrm{s}$ with $\Delta = 0$. In Fig.~\ref{Results}e-f, the red lines display parameter paths $\Delta_\mathrm{max} = \pm 10\pi\ \mathrm{rad}/\mu\mathrm{s}$ and black lines display $\Delta_\mathrm{max} = \pm 4\pi\ \mathrm{rad}/\mu\mathrm{s}$ with $J = 3.74\ \mathrm{rad}/\mu\mathrm{s}$ for all. The dashed lines are obtained from simulation that uses the Lindblad master equation to evolve the qutrit, post-select, and calculate the transition probabilities. We observe that $\langle e^{-\beta W}\rangle \simeq 1$ for paths that are confined to the regions of effective PT symmetry, where $\Delta_\mathrm{max}=0$. In contrast, the Jarzynski equality is violated for paths with $\Delta_\mathrm{max} \neq 0$, and in agreement with the simulation. 

\begin{figure} [t]
    \centering
    \includegraphics[width=1\linewidth]{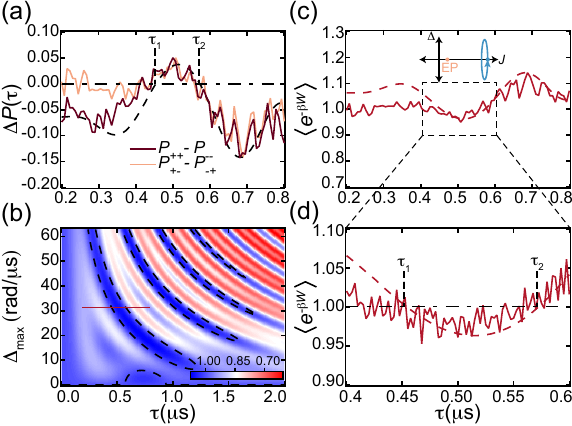}
    %\internallinenumbers
    \caption{{\bf Emergent parity-time symmetry and Jarzynski equality}. For the $\Delta_2(t)$ cyclic path with zero average detuning, the Floquet Hamiltonian always has an emergent $\mathcal{PT}$ symmetry. (a) Yet, the transition probability asymmetries $\Delta P(\tau)=P_{++}-P_{--}$ (brown) and $\Delta P(\tau)=P_{+-}-P_{-+}$ (orange) are generally nonzero except at loop times $\tau_1=0.455\ \mu\mathrm{s}$  and $\tau_2=0.572\ \mu\mathrm{s}$. (b) Simulation of exponentiated-work average \eqref{eq:JE1} while sweeping $\Delta_\mathrm{max}$ and loop time $\tau$. Black dashed contours represent $\langle e^{-\beta W}\rangle=1$ and correspond to cases where the Floquet energy operator $H^F$ aligns with the initial energy operator $H(0)$. Red solid line: experimental parameter space. (c) Measured exponentiated-work average for the $\Delta_2(t)$ path with $\Delta_\mathrm{max} = 10\pi\ \mathrm{rad}/\mu\mathrm{s}$ (solid line) shows the Jarzynski equality is satisfied at loop times $\tau_1,\tau_2$. (d) A higher-resolution in loop-time measurement of (c) for $\tau$ ranging from $0.4\ \mu\mathrm{s}$ to $0.6\ \mu\mathrm{s}$. Dashed lines in (a),(c),(d): simulation.}
    \label{exp_emergent}
\end{figure}

Lastly, we investigate a case where $H_\mathrm{eff}(t)$ has no explicit $\mathcal{PT}$ symmetry, and yet its Floquet counterpart $H^F_{\mathrm{eff}}$ is parity-time symmetric. In this case, the Jarzynski equality is satisfied only at specific loop times: times where the Floquet energy operator $H^F$ aligns with the system's initial energy operator $H(0)=J_{\max}\sigma_x$. We introduce a new parameter path, Eq.~(\ref{Delta2vary}), which obeys $\Delta_2(t)=-\Delta_2(\tau-t)$. As a consequence of the zero average detuning, the corresponding Floquet Hamiltonian has an emergent parity-time symmetry, i.e. $H^F_{\mathrm{eff}}$ eigenvalues are always real or complex conjugates. Figure~\ref{exp_emergent}a shows that the measured probability asymmetry $\Delta P(\tau)\equiv P_{++}(\tau)-P_{--}(\tau)$ (brown), or equivalently, $\Delta P(\tau)=P_{+-}(\tau)-P_{-+}(\tau)$ (orange), is generally nonzero. However, the symmetry under the eigenstate-label exchange $+x\leftrightarrow -x$, indicated by $\Delta P(\tau)=0$, is recovered at loop times $\tau_1=0.455$ $\mu\mathrm{s}$ and $\tau_2=0.572$ $\mu\mathrm{s}$. The corresponding simulated exponentiated-work average shows that although generally violated, the Jarzynski equality is satisfied along black dashed contours (Fig.~\ref{exp_emergent}b). 
%Exactly along these contours, the Floquet energy operator $H^F=(H^F_{\mathrm{eff}}+H^{F\dagger}_{\mathrm{eff}})/2\propto H(0)$. 
These contours intersect with the experimentally investigated region at $\Delta_{\max}=10\pi$ rad/$\mu\mathrm{s}$ (red solid line). In general, the experimentally measured $\langle e^{-\beta W}\rangle(\tau)$ is not equal to unity (Fig.~\ref{exp_emergent}c). However, at loop times $\tau_1$ and $\tau_2$ the two equalities $\langle e^{-\beta W}\rangle=1$ and $\Delta P=0$ are satisfied simultaneously. A higher-resolution measurement of exponentiated-work in a smaller loop-time window shows this effect clearly (Fig.~\ref{exp_emergent}d). 

%This case is unique because, like the non-zero $\Delta$ cases, there is no antilinear symmetry throughout the loop. However, the Floquet Hamiltonian has emergent antilinear symmetry at certain loop times. This results in symmetry of the transition probabilities and verification of the Jarzynski equality at those loop times. Fig.~\ref{exp_emergent}a-b show that the symmetry in the eigenstates' transition probabilities recover at 0.455 and 0.572 $\mu\mathrm{s}$. In Fig.~\ref{exp_emergent}c we show the corresponding $\langle e^{-\beta W}\rangle$ for $J = 5.74\ \mathrm{rad}/ \mu\mathrm{s}$ and $\Delta_\mathrm{max} = 10\pi\ \mathrm{rad}/\mu\mathrm{s}$. The parameter path is shown in the upper inset. The lower inset is a higher resolution run from 0.4 to 0.6 $\mu$s to show the crossing over $e^{-\beta\Delta F} = 1$ more clearly. To show that this effect of recovering antilinear symmetry extends to more parameter values, we plot in Fig.~\ref{exp_emergent}d the simulation of \eqref{LHS} while sweeping the detuning $\Delta_\mathrm{max}$ and loop time $\tau$ with constant $J = 5.74\ \mathrm{rad}/ \mu\mathrm{s}$.\\

%===================== FIG 5 ========================================%
%===================== FIG 5 ========================================%
%\normalem

%=========================================================================================%
%\resetlinenumber[219]
\noindent{\bf Discussion and Outlook.} Non-Hermitian Hamiltonians with real spectra~\cite{Bend98}, first realized in open classical systems~\cite{guo09,zdem19,Miri19}, have recently materialized in the quantum domain~\cite{nagh19,Li2019,Klauck2019,Wu2019,Ding2021,Mara2022}. On top of their role in dynamics, their real eigenvalues are often mistaken for allowed energies of a quantum system~\cite{Def15,Gardas2016,Zeng2017,Bobo2018,Boboa2018,Zhou2021}, with implications to thermodynamics. Although the two share conceptual roots, the thermodynamics of non-Hermitian systems remains an open challenge. A consistent formulation of its first law requires distinguishing the Hermitian part $H$ that gives allowed energies~\cite{nagh20} from the non-Hermitian Hamiltonian $H_\mathrm{eff}=H+i\Gamma$ that governs the temporal dynamics. Using the same distinction we have verified a fluctuation theorem for exponentiated work, i.e. Jarzynski equality \eqref{eq:JE1} for cyclic variations of $H_\mathrm{eff}(t)$ that include parameter regions with complex eigenvalues. 

The Jarzynski equality, rigorously tested in the classical domain~\cite{Liphardt2002,Saira2012}, trivially extends to isolated quantum systems ($Q=0$) by equating the work distribution $P(W)$ with the TPM protocol that, technically, generates the distribution $P(\Delta U)$ of internal-energy changes~\cite{Hub08}. It also holds in decohering quantum systems~\cite{Smi18}, but the equality's validity in driven qubits with dissipation, continuous monitoring, or feedback, as tested with the TPM protocol, is disputed~\cite{Hekking2013,Pek15}. In such settings, the tests of Jarzynski equality require modifications that reflect the energetic cost of information or, equivalently, new dynamics that encode the monitoring, feedback, and measurement processes~\cite{Toyabe2010,Gong2016,Masuyama2018,Nagh18,song21,Lin22}. Our results show that the coherent qubit dynamics of non-Hermitian Hamiltonians is a new class where the Jarzynski equality for cyclic variations is preserved when two symmetry considerations are met. First, the presence of explicit or emergent parity-time symmetry ensures that the qubit has symmetrical amplification or decay rates. Second, energetic changes of the qubit (sometimes termed quantum heat~\cite{Elo17}) vanish when the measurement basis of $H(0)=H(\tau)$ ($\ket{\pm x}$, in the present case) align with the equivalent Floquet energy operator $H^F$ basis. %In this case these measurements are non-invasive. (Due to the absence of quantum jumps, there is no classical heat.) 
With our symmetry-governed, consistent formulation of the second law of thermodynamics, we anticipate new opportunities in quantum, non-equilibrium thermodynamics through non-Hermitian models.\\

\noindent{\textbf{Methods:}}\\
\textbf{Analytical Results.} For a qubit with internal energy operator $H(0)=H(\tau)=J_{\max}\sigma_x$, the exponentiated-work average \eqref{LHS} is given by 
\begin{linenomath}
\begin{align}
    \langle e^{-\beta W}\rangle&=e^{-2\beta J_{\max}}P_{+-}(\tau)P_{-x}+ e^{+2\beta J_{\max}}P_{-+}(\tau)P_{+x}\nonumber \\
    &+ P_{++}(\tau)P_{+x}+P_{--}(\tau)P_{-x}. 
\end{align}
\end{linenomath}
It is easy to verify that the right-hand side is equal to unity when the initial density matrix is thermal, i.e. $P_{\pm x}=\exp(\mp\beta J_{\max})/2\cosh(\beta J_{\max})$, and the transition probabilities are symmetric under the eigenstate-label exchange $+x\leftrightarrow -x$ (Fig.~\ref{Probvstime}a). The exchange-symmetry constraint on the $P_{ij}(\tau)$ holds provided the elements of the time-evolution matrix $G(\tau)$ satisfy $|G_{++}/G_{-+}|=|G_{--}/G_{+-}|$. 
% \begin{align}
% \label{eq:gratio}
%     \frac{G_{++}(\tau)}{G_{-+}(\tau)}=e^{i\phi}\frac{G_{--}(\tau)}{G_{+-}(\tau)}
% \end{align}
% where $\phi$ is an arbitrary phase factor that can be eliminated by redefining $\ket{-x}$ as $\ket{-x}=e^{-i\phi/2}(\ket{f}-\ket{e})/\sqrt{2}$. 
By expressing the traceless non-Hermitian Floquet Hamiltonian as $H^F_\mathrm{eff}=h_x\sigma_x+h_y\sigma_y+h_z\sigma_z$, the exchange-symmetry constraint can be written as
\begin{linenomath}
\begin{align}
\label{eq:cs}
\left|\frac{\mathcal{C}-ih_x\mathcal{S}}{(h_y+ih_z)\mathcal{S}}\right|=\left|\frac{\mathcal{C}+ih_x\mathcal{S}}{(h_y-ih_z)\mathcal{S}}\right|
\end{align}
\end{linenomath}

where $\mathcal{C}=\cos(\tau|{\bf h}|)$, $|{\bf h}|\equiv(h_x^2+h_y^2+h_z^2)^{1/2}$, and $\mathcal{S}=\sin(\tau|{\bf h}|)/|{\bf h}|$. The terms $\mathcal{C, S}$ are real if and only if $|{\bf h}|$ is real or purely imaginary, which means the Floquet Hamiltonian $H^F_\mathrm{eff}$, with eigenvalues $\pm|{\bf h}|$, has an explicit or emergent parity-time (antilinear) symmetry~\cite{Harter2020}. Eq.~(\ref{eq:cs}) further requires that $h_x\in\mathbb{R}$ and $h_y,h_z$ are purely imaginary. Thus, the non-Hermitian, parity-time symmetric, Floquet Hamiltonian $H^F_\mathrm{eff}=H^F+i\Gamma^F$ is further constrained to an internal energy operator $H^F=h_x\sigma_x$ that is aligned with the system's initial energy operator $H(0)$. Note that the mere requirement of parity-time symmetry allows for $h_x\in\mathbb{R}$ and a complex $h_y=h_z^*\in\mathbb{C}$. However, in such cases, $H^F=h_x\sigma_x+\Re h_y(\sigma_y+\sigma_z)$ is not aligned with $H(0)$, and the constraint that guarantees exchange symmetry for probabilities, Eq.\eqref{eq:cs}, is not fulfilled. Thus, Jarzynski equality requires a parity-time symmetric $H^F_\mathrm{eff}$ with its Floquet energy operator proportional to the system's initial energy operator.  

\noindent\textbf{Experimental setup}. The experimental setup comprises a superconducting circuit that was fabricated and provided by the Superconducting Qubits at Lincoln Laboratory (SQUILL) Foundry at MIT Lincoln Laboratory. The experiments utilize a sub-portion of a multi-qubit chip with relevant components consisting of a tunable transmon qubit with maximum frequency $\omega_{ge}/2\pi = 4.373\ \mathrm{GHz}$, dispersively coupled to a readout resonator at coupling rate $g/2\pi = 33\ \mathrm{MHz}$ and linewidth $\kappa/2\pi = 246\ \mathrm{kHz}$, qubit drive line, and an off-chip coupling line. A solenoid coil fixed to the package allows control of the global flux through the transmon SQUID loop, with bias current filtered at the 4K stage with a low pass filter (QDevil Q015 QFilter). The qubit is operated at $\omega_{ge}/2\pi = 4.25\ \mathrm{GHz}$ and resonator frequency $\omega_{r}/2\pi = 6.88865\ \mathrm{GHz}$. To realize the non-Hermiticity, an off-chip coaxial filter is coupled to the qubit to enhance the $\ket{e}$ decay rate to $\gamma_e=1.57\ \mu\mathrm{s}^{-1}$. The readout signal probes the resonator via a common bus line and is amplified by a Josephson parametric amplifier (BBN-PS2-JPA-DEVICE-QEC) operating with $\sim 15$ dB of gain.

%This experiment utilizes a superconducting transmon circuit \cite{koch07} fabricated by the SQUILL MIT Lincoln Labs foundry\serracom{(*Kater should we cite this/what is the official way to acknowledge this?)}. It is an aluminum design deposited on a silicon chip. There is a coil attached to the package that controls the global flux to the chip via DC lines, which allows for a flux tunable qubit. We utilize the lower three energy levels of the transmon and operate it at $\omega_{ge}/2\pi = 4.25\ \mathrm{GHz}$. The qubit is coupled to a readout resonator with coupling rate of $g/2\pi = 48\ \mathrm{MHz}$ and frequency $\omega_{r}/2\pi = 6.88865\ \mathrm{GHz}$. We have a decay channel with decay rate $1.496\ \mu\mathrm{s}^{-1}$ weakly coupled to the qubit at resonance $4.25\ \mathrm{GHz}$. This allows us to tune the dissipation rate of the excited state to satisfy $\gamma_e \gg \gamma_f$. At the $4\mathrm{K}$ stage we have a low-pass filter on the DC lines with cut off frequency of $65\ \mathrm{kHz}$ from QDevil \serracom{(*part number)}. To achieve higher fidelity readout we have a Josephson parametric amplifier (JPA) from Raytheon BBN \serracom{(*part number)} connected to the output line, providing $15\ \mathrm{dB}$ of gain.

\noindent{\textbf{Simulations}}. The evolution of the three-level system can be solved using Lindblad equation
\begin{equation}
       \frac{ \partial \rho_3(t)}{\partial t} = - i [H_{c}(t),\rho_3(t)]+ \sum_{i=1}^{4} L_i \rho_3(t) L_i^{\dagger}- \frac{1}{2}\{L_i^{\dagger}L_i,\rho_3(t)\}.
       \label{lindbladeq}
\end{equation}
Here $H_c= J(\ketbra{e}{f}+\ketbra{f}{e}) -\Delta/2(\ketbra{e}{e}-\ketbra{f}{f})$ and the four dissipators $L_i$ include two radiative decay operators $\sqrt{\gamma_e }\ketbra{g}{e}$ and $\sqrt{\gamma_f} \ketbra{e}{f}$, and two dephasing operators $\sqrt{\gamma_{2e }/2} \ketbra{e}{e}$ and $\sqrt{\gamma_{2f }/2} \ketbra{f}{f}$. The decay and dephasing rates are: $\gamma_e = 1.57\ \mu\mathrm{s}^{-1}$, $\gamma_f = 0.21\ \mu\mathrm{s}^{-1}$, $\gamma_{2e } = 1.631\ \mu\mathrm{s}^{-1}$, and $\gamma_{2f } = 0.584\ \mu\mathrm{s}^{-1}$. Eq.\eqref{lindbladeq} is solved in MATLAB using the Runge-Kutta method to obtain $\rho_3(\tau)$ with suitable initial conditions and thereby calculate each transition probability $P_{ij}(\tau)$. For $\gamma_f,\gamma_{2e},\gamma_{2f}\ll\gamma_e$, the Lindblad results for the $\{\ket{e},\ket{f}\}$ manifold are identical to those obtained from the non-Hermitian Hamiltonian~\cite{Varma2023},
\begin{linenomath}
\begin{align}
    \rho_2(\tau)=\frac{G(\tau)\rho_2(0)G^\dagger(\tau)}{\mathrm{Tr}[\rho(0)G^\dagger(\tau)G(\tau)]}.
\end{align}
\end{linenomath}

\noindent{\textbf{Post-selection and error analysis}}.
For the experimental data, we employ post-selection by normalizing the state readouts to the population within the $\{|e\rangle,|f\rangle\}$ manifold of states, the resulting evolution can be described by the non-Hermitian Hamiltonian \eqref{H}.
%\begin{equation}
%H_\mathrm{eff}= J(|e\rangle\langle f|+|f\rangle\langle e|) -(i \gamma_e -\Delta)|e\rangle\langle e|
%\end{equation}
%where the experimental sequences with jumps to the state $|g\rangle$ are discarded. 
For each transition probability, we repeat the experiment a total of 8000 times, yet through post-selection up to $\sim65\%$ of the data is discarded. The statistical (trinomial) error associated with the state readout is typically less than $0.016$ for the transition probabilities and less than $0.012$ for the exponentiated work. Remnant, point-to-point fluctuations are likely due to residual low-frequency (1/f) fluctuations in the experimental setup. For Fig.~\ref{exp_emergent}a,c-d we utilized 24,000 experimental repetitions per point.

\noindent{\textbf{Acknowledgements}}\\
We thank A. Auff\`eves and E. Lutz for helpful comments. This research was supported by NSF Grant No. PHY-1752844 (CAREER), the Air Force Office of Scientific Research (AFOSR) Multidisciplinary University Research Initiative (MURI) Award on Programmable systems with non-Hermitian quantum dynamics (Grant No. FA9550-21-1- 0202), the John Templeton Foundation, Grant No. 61835, ONR Grant No. N00014- 21-1-2630, and the Institute of Materials Science and Engineering at Washington University. Devices were fabricated and provided by the Superconducting Qubits at Lincoln Laboratory (SQUILL) Foundry at MIT Lincoln Laboratory, with funding from the Laboratory for Physical Sciences (LPS) Qubit Collaboratory.

\end{document}